\documentclass[aps,prd,groupedaddress,graphicx,nofootinbib,onecolumn]{revtex4}
\usepackage{amssymb,graphics,graphicx,color}

\begin{document}

\title{Primordial Black Holes from Passive Density Fluctuations}

\author{Chia-Min Lin$^1$}\email{cmlin@boar.kobe-u.ac.jp}
\author{Kin-Wang Ng$^2$}\email{nkw@phys.sinica.edu.tw}
\affiliation{$^1$Department of Physics, Kobe University, Kobe 657-8501, Japan}
\affiliation{$^2$Institute of Physics \& Institute of Astronomy and Astrophysics,
Academia Sinica, Taipei 11529, Taiwan}

\date{Draft \today}

\begin{abstract}
In this paper, we show that if passive fluctuations are considered, primordial black holes (PBHs) can be easily produced in the framework of single-field, slow-roll inflation models. The formation of PBHs is due to the blue spectrum of passive fluctuations and an enhancement of the spectral range which exits horizon near the end of inflation. Therefore the PBHs are light with masses $\lesssim 10^{15}\mbox{g}$ depending on the number of e-folds when the scale of our observable universe leaves horizon. These PBHs are likely to have evaporated and cannot be a candidate for dark matter but they may still affect the early universe.
\end{abstract}
\maketitle

\section{Introduction}

Inflation \cite{Lyth:2009zz} is becoming a standard model for the very early universe.
The inflationary scenario, in which the present
universe is only a small local patch of a causally connected
region at early times which underwent an exponential expansion
driven by the inflaton potential, is generally accepted for
explaining the observed spatially flat and homogeneous universe.
In addition, its quantum fluctuations during inflation give rise
to primordial Gaussian matter density fluctuations with a nearly
scale-invariant power spectrum, which is consistent with recent
astrophysical and cosmological observations such as structure
formation and cosmic microwave background
anisotropies~\cite{texas04}.

Although the simplest single-field, slow-roll inflation model
works well, some basic questions have yet to be answered. What is
the origin of the inflaton potential? Do classical matter density
imhomogenities that we observe today genuinely come from quantum
fluctuations of the inflaton? Are the observed matter density
fluctuations truly Gaussian? How robust are the predictions for a
subdominant contribution of tensor modes to the metric
fluctuations, a slightly broken scale invariance, and a negligible
running spectral index of the power spectrum? Future cosmic
microwave background measurements and mega-scale mappings of the
large scale structure will definitely answer some of these
questions or perhaps pose a challenge to the standard inflation scenario.

There has been a lot of studies on inflationary models that go
beyond the simplest single-field, slow-roll inflation. A class of
models has considered a new source for generating inflaton
fluctuations (so-called passive density fluctuations) during inflation
through a direct or gravitational coupling between the inflaton and other quantum
fields. This leads to very interesting results such as the
so-called warm inflation~\cite{berera}, the suppression of
large-scale density fluctuations~\cite{wu}, possible constraints on the
duration of inflationary expansion~\cite{wood}. the bursts of particle
production  that result in infra-red cascading~\cite{barnaby},
the trapped inflation in which the inflaton rolls slowly down a
steep potential by dumping its kinetic energy into light particles at the trapping
points along the inflaton trajectory~\cite{green,lee}, and electromagnetic dissipation
in natural inflation~\cite{sorbo,Barnaby:2010vf, Barnaby:2011vw,Meerburg:2012id}.

In all of these papers, essentially, the generation of passive density fluctuations is originated from quantum
fluctuations in the back reaction of the couplings to the inflaton perturbation. The nature of passive fluctuations is usually non-Gaussian and non-scale-invariant. A particular feature is that
the power spectrum of the passive fluctuations can be very blue~\cite{wood,lee,sorbo,Barnaby:2010vf, Barnaby:2011vw,Meerburg:2012id}. These passive fluctuations cannot dominate the primordial density perturbation at large scales as confirmed by cosmological observations such as cosmic microwave background (CMB) anisotropies and the formation of large scale structures. However, depending on individual models, their significant contribution to the non-Gaussianity is still possible. This will be tested soon in the
Planck CMB mission and in future large-scale-structure surveys. In this paper, we point out that the passive fluctuations with a blue spectrum can dominate the primordial density perturbation in the very small scales,
seeding the formation of primordial black holes (PBHs) in the radiation-domination era after inflation.

\section{Passive Density Fluctuations During Inflation}

Let us consider a slow-rolling inflaton $\phi$ coupled to
a certain quantum field $\chi$. The Lagrangian that is relevant to
$\phi$ is given by
\begin{equation}
L=\frac{1}{2}g^{\mu\nu}\partial_{\mu}\phi\,\partial_{\nu}\phi
             -V(\phi) + L_I(\phi,\chi),
\end{equation}
where $V$ is the inflaton potential, $L_I$ is the interaction term,
and the metric is
\begin{equation}
ds^{2}=dt^2-a^2(t) d{\bf x}^2.
\end{equation}
The equations of motion for the mean fields are then given by
\begin{eqnarray}
&&H^2\equiv \left(\frac{\dot a}{a}\right)^2 =\frac{8\pi}{3M_P^2}
\left({1\over2}{\dot{\phi}}^2+ V +\rho_\chi\right), \\
&&\ddot{\phi}+3H\dot{\phi}+ V'=\frac{\partial L_I}{\partial\phi},
\label{backreaction}
\end{eqnarray}
where $\rho_\chi$ is the energy density of $\chi$ particles,
the dot and the prime denote differentiating with respect to
time and $\phi$ respectively, and $M_P=2.4 \times 10^{18}\mbox{GeV}$
is the reduced Planck mass. In Eq.~(\ref{backreaction}), the righthand
side of the equation is the back reaction of the interaction to
the inflaton mean field. The back reaction arises due to copious
production of $\chi$ quanta during inflation.

In addition, the fluctuations of $\phi$ satisfy
\begin{equation}
\ddot{\delta\phi}+3H\dot{\delta\phi} -\frac{\nabla^{2}}{a^2}{\delta\phi}+ V''{\delta\phi}=
\delta\left(\frac{\partial L_I}{\partial\phi}\right).
\label{flucteqn}
\end{equation}
The homogeneous solution of this fluctuation equation
gives rise to the standard primordial density fluctuations. Here we call them
as active fluctuations and denote their power spectrum by $P_a$.
The righthand side of Eq.~(\ref{flucteqn}), which comes from the fluctuations of
the backreaction, acts as a source for generating additional fluctuations of $\phi$.
The particular solution of Eq.~(\ref{flucteqn}) with the source term is referred
as passive fluctuations and the power spectrum is denoted by $P_p$. Hence, the total
power spectrum is given by the contributions from both active fluctuations and passive fluctuations as
\begin{equation}
P=P_a+P_p=\left(\frac{H}{\dot\phi}\right)^2 \langle |\delta\phi_k|^2 \rangle,
\end{equation}
where $\delta\phi_k$ is the Fourier mode of $\delta\phi$.

\section{Primordial Black Holes}

For single-field slow-roll inflation, the spectrum from active fluctuations is given by
\begin{equation}
P_a=\frac{1}{24\pi^2 M_P^4}\frac{V}{\epsilon},
\label{eq2}
\end{equation}
where the slow-roll parameter is
\begin{equation}
\epsilon \equiv \frac{M_P^2}{2} \left( \frac{V^\prime}{V} \right).
\label{epsilon}
\end{equation}

In a given region with radius $r$, the criteria of black hole formation is given by
\begin{equation}
\frac{2G\delta M}{r}=\frac{2 \delta M}{rM_P^2}>1
\end{equation}
where $G$ is Newton's constant and $\delta M$ is the mass inside the region $r$.\footnote{Interestingly this relation can be found from Newtonian physics by requiring the escape velocity to be larger than the speed of light.} The condition can be expressed by using the energy density $\delta M \sim \delta \rho r^3$ as
\begin{equation}
\delta\rho \gtrsim \frac{M_P^2}{r^2}.
\label{eq3}
\end{equation}

The primordial density perturbation can be imagined as density fluctuation between different Hubble patches of the universe (so called `separate universes' \cite{Wands:2000dp}) with radius $r \sim 1/H$. Namely, in each patch of the universe, the energy density is regarded as homogeneous but each patch has different value of energy density. Therefore Eq.~(\ref{eq3}) becomes $\delta \rho \gtrsim M_P^2 H^2 \sim \rho$ where in the second equality Friedmann equation ($\rho \sim 3H^2 M_P^2$) is used and $\rho$ should be regarded as the average of many patches. From this simple estimation, we can naively guess that if $P^{1/2} \sim (\delta \rho / \rho)  \gtrsim \mathcal{O} (1)$, the whole patch of the separate universe will colapse into a black hole. The argument here is heuristic one. For those who concern about the gauge dependence of density fluctuation, our argument has assumed a spatially flat gauge in which we choose a slice with zero curvature perturbation and hence the relevant quantity is density fluctuation.
More rigorous calculation for the condition of primordial black hole formation is given by Refs.~\cite{Carr:1974nx, Yokoyama:1995ex} as
\begin{equation}
\frac{1}{3} < \delta \equiv \frac{\delta \rho}{\rho} <1.
\label{eq5}
\end{equation}
The upper bound is to avoid formation of a separate closed universe.
If the spectrum is $P^{1/2} \sim {\cal O}(10^{-2})$, PBHs will be copiously produced \cite{Carr:1974nx}.
This is certainly not the range of the matter power spectrum probed by CMB experiments. The range corresponds to the
scale of quantum fluctuations that exits the horizon during inflation at a number of e-folds, $N=60$, before inflation ends. It is because we have $P(N=60)=(5 \times 10^{-5})^2$ from CMB observations.
However, the spectrum can be large near the end of inflation,
\begin{equation}
P(N=0) \sim 10^{-4} \sim 10^{6} \times P(N=60).
\end{equation}
In this case black holes will form soon after inflation when the scale enters the horizon. These black holes are called primordial black holes (PBHs) (see Refs.~\cite{Khlopov:2008qy, Carr:2003bj} for review). The spectrum from active fluctuations is (almost) scale invariant, therefore unless the running spectral index is large, PBHs cannot be formed~\cite{Alabidi:2009bk, Kohri:2007qn}. We can see from Eqs.~(\ref{eq2}) and (\ref{epsilon}) that the slow-roll parameter $\epsilon$ has to decrease toward the end of inflation in order to enhance the (active) spectrum. However, inflation has to end so the usual tendency is an increasing $\epsilon$. This is the reason why it is so difficult to have PBHs (for single-field slow-roll inflation with active fluctuations) \cite{Kohri:2007gq, Bugaev:2008bi, Bugaev:2008gw, Drees:2011yz}. However, we point out for the first time that it is very natural and easy to have PBHs formed even in the case of single-field slow-roll inflation if we consider passive fluctuations.

\section{Inflation Models and Passive Power Spectra}
In this section, we will discuss the passive fluctuations in different inflation models
and explore the possibility of forming primordial black holes from the passive fluctuations.

\subsection{Axion Inflation}
\label{sec1}
Let us consider the case that the inflaton $\phi$ is a pseudo Nambu-Goldstone boson with a typical potential after the shift symmetry is broken,
\begin{equation}
V(\phi)=\Lambda^4 [1-\cos(\phi/f)],
\label{axion}
\end{equation}
where $\Lambda$ is a mass scale and $f$ is the axion decay constant. And the coupling to a gauge field is
\begin{equation}
L_I = -\frac{\alpha}{4f}\phi F^{\mu\nu}\widetilde{F}_{\mu\nu}.
\end{equation}
This coupling is natural in the sense that there is no symmetry to forbid it.
It is found~\cite{Barnaby:2010vf, Barnaby:2011vw} ( and been further analysised in Ref.~\cite{Meerburg:2012id} ) that the inverse decay of the gauge field will enhance the fluctuation of the inflaton field. The spectrum is given by
\begin{equation}
P=P_a \left( 1+ 7.5 \times 10^{-5} P_a \frac{e^{4 \pi \xi}}{\xi^6} \right),
\label{eq10}
\end{equation}
where the second term corresponds to the passive spectrum $P_p$ and
\begin{equation}
\xi \equiv \frac{\dot{\phi}\alpha}{2Hf}=\sqrt{\frac{\epsilon}{2}}\frac{\alpha M_p}{f}.
\label{eq12}
\end{equation}

\begin{figure}[t]
  \centering
\includegraphics[width=0.6\textwidth]{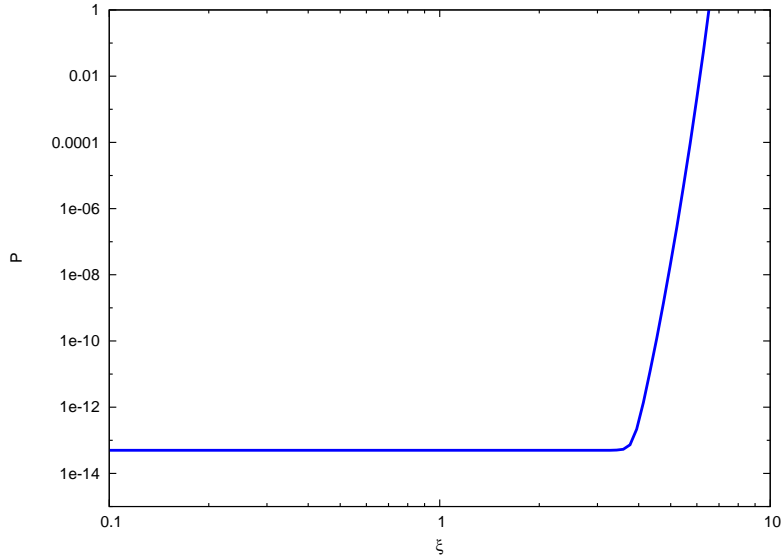}
  \caption{Power spectrum produced near the end of inflation. If $\xi \sim 6$, primordial black hole is formed.}
  \label{fig1}
\end{figure}

At CMB scale ($N=60$), large $\xi$ would produce large non-Gaussainity according to Eq.~(\ref{eq10}). Therefore, there is an upper bound from CMB data which is roughly $\xi \lesssim 3$  \cite{Barnaby:2010vf, Meerburg:2012id}. However, $\xi$ becomes larger after CMB scale exits horizon. For example, if we approximate Eq.~(\ref{axion}) by a quadratic potential $V=m^2 \phi^2/2$ with $m=\Lambda^2 /f$ which is a good approximation when $\phi \ll f$,\footnote{Actually this will result in $f>M_P$ which is not phenomenologically desirable, but it can be evaded for example if we consider the potential as an effective potential of N-flation \cite{Dimopoulos:2005ac}. See \cite{sorbo} for more examples in multi-field or extra-dimensional models where it is very natural to have $f<M_P$ and $\alpha/f \gg 1/M_P$.}we would have the slow-roll parameter $\epsilon \sim 0.01$ at $N=60$ (CMB scale) and $\epsilon \sim 1$ at $N=0$. This means that $\xi$ grows ten times larger according to Eq.~(\ref{eq12}). The total spectrum (near the end of inflation) is plotted in Fig.~\ref{fig1} by using Eq.~(\ref{eq10}). We can see from the figure that when $\xi(N=0) \sim 6$, Eq.~(\ref{eq5}) is satisfied and PBHs would form. This would correspond to $\xi(N=60) \sim 0.6$ (noting that this can be achieved by $\alpha \sim 1$ if $f<M_P$), which is well below the constraint ($\xi \lesssim 3$) from Refs.~\cite{Barnaby:2010vf, Meerburg:2012id}. This means that we do not have large non-Gaussianity but we have PBHs. Actually the abundance of produced PBHs may put an upper bound on the non-Gaussianity generated in this model.

One may concern about the effect of backreaction if we have $\xi \sim 6$ near the end of inflation. First of all, this effect is milder in our case than the case where large non-Gaussianity is generated at CMB scale. In both cases, $\xi$ is expected to grow toward the end of inflation and we consider a smaller value of $\xi$ at $N \sim 60$. It has been shown in Ref.~\cite{Meerburg:2012id} that the  backreaction to the homogeneous inflaton field due to particle production slows down the inflaton velocity and prolongs the end of inflation (instead of shortening the period of inflation). However, this does not change our results since the power spectrum depends only on $\xi$, which in turn depends on the slow-roll parameter $\epsilon$, and the end of inflation is determined by $\epsilon=1$.

\subsection{Inflation with Trapping}

Consider an interaction between $\phi$ and massless scalar $\chi_i$ of the type,
\begin{equation}
L_I= g^2\sum_i (\phi-\phi_i)^2\chi_i^2,
\end{equation}
where $g$ is a coupling constant and $\phi_i$ is a
constant field value. When $\phi$ rolls down to each trapping point at $\phi_i$
along the inflaton trajectory, the $\chi_i$ particles become instantaneously
massless and are produced with a number density that increases
with $\phi$'s velocity. As $\phi$ dumps its
kinetic energy into the $\chi_i$ particles, it is slowed down and
the produced $\chi_i$ particles are diluted due to the inflationary
expansion.

It is useful to define a time scale, $\Delta t \equiv 1/\sqrt{g{\dot\phi}}$.
Then, We have
\begin{equation}
H\Delta t=\left(\frac{2\pi}{g}\right)^{1\over2} P_a^{1\over4}.
\end{equation}
For $H\Delta t < 1$ or $g^2>10^{-7}$,
it was shown~\cite{barnaby,green} that bursts of $\chi_i$ particle production takes place
in a time scale, $\Delta t$, and the number density of
the $\chi_i$ particles produced is given by $n_{\chi_i} \simeq 1/\Delta t^3$.
This leads to the backreaction of the $\chi_i$ particle production to
the inflaton field in Eq.~(\ref{backreaction}) and the righthand side of Eq.~(\ref{flucteqn}) is
particle number density fluctuations in the process of particle production.
Furthermore, for closely spaced trapping points with equal spacing $\Gamma$
along the inflaton trajectory, trapped inflation occurs even on a
potential which is too steep for slow-roll inflation, provided that
$H\Gamma/{\dot\phi}\ll 1$~\cite{green}.

In the case for $H\Delta t \simeq 1$ (i.e. $g^2\simeq 10^{-7}$) and $\Gamma\simeq {\dot\phi}\Delta t$,
it was found~\cite{lee} that $P_p$ has a blue power spectrum. Extrapolating the result obtained
in Ref.~\cite{lee}, we find that the total power spectrum in this case is approximately given by
\begin{equation}
P=P_a \left[ 1+0.28 g^2\left(\frac{k}{k_{CMB}}\right)^2\right],
\end{equation}
where the pivotal scale $k_{CMB}=0.002\mbox{Mpc}^{-1}$, if inflation lasts for about $60$ e-folds.
This reproduces the power spectrum in Figure~$2$ of Ref.~\cite{lee} for $k/k_{CMB}<300$.
Therefore, $P$ can easily reach unity for small scales that exit horizon before the end of inflation. Here we have assumed that the scale dependence does not change dramatically during inflation when doing the extrapolation. This is plausible since inflation is (quasi) de Sitter phase and hence no extra physical effect is expected to change this scaling behavior during inflation. This blue-tilted behaviour is reasonable because a certain $k$-mode is being kicked constantly by the trapping fields all along the inflaton trajectory and the kicks result in accumulative growth of the fluctutations (see Eq.~(54) of Ref.~\cite{lee}). The growth is bigger for higher $k$ modes since they leave the horizon later in time.

\subsection{Quantum Stress Tensor Fluctuations}

The last model does not involve a direct non-gravitational interaction between
inflaton and other quantum fields as discussed above. The passive fluctuations
come from gravitational effects and are therefore intrinsic and model independent.
In Ref.~\cite{wood}, the authors considered the effect of quantum stress tensor fluctuations
of conformal fields such as conformally coupled massless scalars and electromagnetic fields,
denoted by $\delta T_{\mu\nu}$,
in de Sitter spacetime upon the expansion of a congruence of timelike geodesics.
They found that the quantum stress tensor fluctuations induce a source term on the righthand side
of Eq.~(\ref{flucteqn}), given by
\begin{equation}
\frac{V'}{3H} \delta\theta,
\end{equation}
where $\theta$ is the expansion of the congruence, related to the covariant derivative of the
fluid four-velocity by $\theta=u^\mu_{;\mu}$. In the presence of quantum stress tensor fluctuations,
they found that
\begin{equation}
\delta\theta =-a^{-2}(t)\int_{t_0}^t dt' a^2(t') \delta R_{\mu\nu} u^\mu u\nu,
\end{equation}
where $t_0$ is when inflation begins and $\delta R_{\mu\nu}$ is the Ricci tensor
fluctuations induced by $\delta T_{\mu\nu}$, from which the passive power spectrum
was obtained as
\begin{equation}
P_p=10^{-49} S \left(\frac{E_R}{10^{12}\mbox{GeV}}\right)^3 \left(\frac{k}{3 \mbox{Mpc}^{-1}}\right),
\end{equation}
where $S$ is the total expansion factor during inflation and $E_R\sim V^{1/4}$ is the reheating
energy scale at the end of inflation. For example, if we consider the scale
leaving horizon near the end of inflation at which PBHs form, we would have $k \sim e^{60} k_{CMB}$.
Assuming that $S>e^{60}$ and $E_R=10^{12}\mbox{GeV}$, we can easily obtain a condition that $P_p\sim 1$.

\section{Primordial Black Holes as Dark Matter?}

When a PBH forms, roughly speaking the whole horizon collapses into a black hole and therefore the mass is given by the horizon mass,
\begin{equation}
M \simeq \frac{4 \pi}{3} \rho \left( \frac{1}{H} \right)^3 = 4 \pi \frac{M_P^2}{H} \simeq 0.066 M_{\odot} \left( \frac{T}{\mbox{GeV}} \right)^{-2} \left( \frac{g}{50} \right),
\end{equation}
where the last equality assumes radiation domination (with energy density $\rho_R = (\pi^2/30)g T^4$) when black holes form\footnote{For estimation and simplicity, we assume that reheating happens immediately after inflation and therefore PBHs always form in a radiation-dominated universe.}, $M_{\odot} \sim 10^{33} \mbox{g}$ is the solar mass,  $T$ is the temperature, and $g$ is the statistical degrees of freedom. By using the formula of entropy density, $s \sim g T^3$, the conservation of the total entropy, $S=a^3 s$, implies that $T \propto g^{-1/3} /a$. Thus,
the PBH mass can also be written as a function of comoving wave number $k$ as \cite{Kanazawa:2000ea}
\begin{equation}
M \simeq 6.4 \times 10^{14} M_\odot \left( \frac{g}{50} \right)^{-1/6} \left( \frac{k}{\mbox{Mpc}^{-1}}\right)^{-2}.
\label{eq11}
\end{equation}

For example, if the PBH forms at $T=10^8 \mbox{GeV}$, we have $M \sim 10^{15}\mbox{g}$. A PBH with mass smaller than $10^{15}\mbox{g}$ would have evaporated through Hawking radiation \cite{Hawking:1974rv}. PBHs with larger masses contribute to (or might even dominate) the dark matter density. The mass fraction $\beta \equiv \rho_{BH}/\rho$ of PBHs of mass $M$ is given by~\cite{Green:1997sz}
\begin{equation}
\beta(M)=\int^1_{1/3}\frac{d\delta}{\sqrt{2\pi}\sigma(M)}\exp\left( -\frac{\delta^2}{2 \sigma^2(M)} \right) \simeq \sigma(M)\exp\left( -\frac{1}{18\sigma^2(M)} \right),
\end{equation}
where $\sigma(M)$ is the mass variance at the horizon crossing. The range of integration corresponds to~Eq.~(\ref{eq5}). If PBHs form during the radiation-dominated epoch, $\sigma \simeq  0.65 \delta$. Hence, the density parameter $\Omega_{BH}(M)$ of the PBHs in the present universe is~\cite{Kanazawa:2000ea}
\begin{equation}
\Omega_{BH}(M)h^2 \simeq 2.1 \times 10^8 \beta(M)\left(\frac{T}{\mbox{GeV}}\right),
\end{equation}
where $h$ is the present Hubble constant in units of $100$ km/sec/Mpc. For example, if the PBHs with $M \sim M_{\odot}$ play the role of dark matter in the present universe, i.e. $\Omega_{BH}h^2 \simeq 0.25$, then $\beta \simeq 5 \times 10^{-9}$, $\sigma \simeq 0.06$, and $\delta \simeq 0.092$.
For the scale of PBH formation leaving horizon near the end of inflation, we would have
$k \sim e^{60} k_{CMB} \sim 10^{23} \mbox{Mpc}^{-1}$. By using Eq.~(\ref{eq11}), we have the PBH mass
$M \sim 10^{-32} M_\odot \sim 20\mbox{g}$. These small black holes would have already evaporated and could not be dark matter. We thus conclude that PBHs generated from scales leaving horizon near the end of inflation are unlikely to become dark matter. However, if we consider low-scale inflation in which the CMB scale corresponds to $N \sim 45$, we may have PBHs with $M\sim 10^{15}\mbox{g}$ that are evaporating now. This may explain the observed antiproton fluxes from the BESS experiment~\cite{Kanazawa:2000ea, Yoshimura:1995sa}. Furthermore, PBHs with $M<10^{15}\mbox{g}$ can have other cosmological implications. For examples, they could affect baryogenesis~\cite{Dolgov:2000ht} and nucleosynthesis~\cite{Kohri:1999ex, Carr:2009jm}, swallow monopoles~\cite{Izawa:1984ww, Stojkovic:2004hz}, destroy domain walls \cite{Stojkovic:2005zh}, and so on (see Ref.~\cite{Carr:2005zd} for more detailed references).

\section{Conclusion}
\label{con}
A lot of effort has been put in studying the passive fluctuations during inflation and their imprints on the CMB and in particular the non-Gaussian features. However,
in this paper we have proposed a novel mechanism that passive fluctuations can produce PBHs with scales exiting horizon near the end of inflation. This would result in light PBHs with mass $M \lesssim 10^{15} \mbox{g}$ that may have interesting cosmological consequences in the early universe. We have discussed three models in which passive fluctuations can easily generate PBHs, noting that the characteristic power spectrum is rather generic, namely, the spectrum is extremely blue at small scales.

\acknowledgments
This work was supported in part by the National Science Council, Taiwan, ROC under Grant No.
NSC98-2112-M-001-009-MY3.

\end{document}